\documentclass[preprint,prb,aps]{revtex4}
\usepackage{color}
\makeatletter

\usepackage{amsmath}
\usepackage{graphicx}
\usepackage{bm}
\usepackage{fancyhdr}
\usepackage{amsmath}
\usepackage{graphicx}
\usepackage{amsfonts}
\usepackage{amsbsy}
\usepackage{amssymb}
\usepackage[mathscr]{eucal}

\usepackage{color}

\graphicspath{{./figures/}}

\usepackage{graphicx}
\usepackage{bm}

\usepackage{amsfonts}
\usepackage{amsbsy}
\usepackage{amssymb}
\usepackage[mathscr]{eucal}

\newcommand{\beq}{\begin{equation}}
\newcommand{\eeq}{\end{equation}}
\newcommand{\ba}{\begin{array}}
\newcommand{\ea}{\end{array}}
\newcommand{\bea}{\begin{eqnarray}}
\newcommand{\eea}{\end{eqnarray}}
\newcommand{\bseq}{\begin{subequations}}
\newcommand{\eseq}{\end{subequations}}

\date{today}

\begin{document}

\title{Quantum threshold reflection of He atom beams from rough surfaces}

\author{G. Rojas-Lorenzo and J. Rubayo-Soneira}
\email{german@instec.cu,jrs@instec.cu}
\affiliation{Instituto Superior de Tecnolog\'ias y Ciencias Aplicadas (InsTEC), Universidad de La
Habana, Ave. Salvador Allende No. 1110,  Quinta de Los Molinos, La Habana 10400, Cuba \\}
\author{S. Miret-Art\'es}
\email{s.miret@iff.csic.es}
\affiliation{Instituto de F\'isica Fundamental, Consejo Superior de Investigaciones
Cient\'ificas, \\ Serrano 123, 28006 Madrid, Spain \\}
\author{E. Pollak}
\email{eli.pollak@weizmann.ac.il}
\affiliation{Chemical and Biological Physics Department, Weizmann Institute of Science, 76100,
Rehovot, Israel}

\date{\today}

\begin{abstract}
	
Quantum reflection of thermal He atoms from various surfaces (glass slide, GaAs wafer, flat and structured Cr) at grazing conditions is studied
within the elastic close-coupling formalism. Comparison with the experimental results of B.S. Zhao et al, Phys. Rev. Lett. {\bf 105}, 133203
(2010) is  quite reasonable but the conclusions of  the present theoretical analysis are different from those discussed in the experimental
work. The universal linear behavior observed in the dependence of the reflection probability on the incident wave vector component perpendicular
to the surface is only valid at small values of the component whereas, at larger values, deviation from the linearity is evident,
approaching a quadratic dependence at higher values. The surface roughness seems to play no important role in this scattering. Moreover,
the claim that one observes a transition from quantum to classical reflection seems to be imprecise.

\end{abstract}

\maketitle

\vspace{2cm}


\newpage

\section{Introduction}

Matter wave diffraction and interferometry are very interesting topics because, unlike optical effects observed by photons, these studies
lead to a better knowledge of the interaction between  particles and surfaces.
This interaction is usually divided  into two regions, short and long range distances. The short distances are dominated by the Pauli and
Coulomb repulsion between electrons of the incoming particle
and the surface and the long distances are governed  mainly by the van der Waals-Casimir attraction. We talk about classical reflection when
particles reach the inner region (turning points) corresponding to the repulsive part of the interaction potential. On the other hand, when the
reflection comes from the long range attractive part, one talks about quantum threshold reflection
to be distinguished from above barrier quantum reflection. Lennard-Jones and
Devonshire \cite{lennard} first recognized this behavior in atom-surface scattering and Kohn \cite{kohn}
showed later on that quantum reflection leads to a zero sticking probability at threshold. He pointed out that this reflection is a quantum
interference process between the incoming and reflected waves. Senn \cite{senn} showed  that, for general one-dimensional forces which
vanish as the coordinate goes to $\pm \infty$,
the reflection probability goes to unity at threshold energies except when the potential supports a zero energy resonance state.
The reflection coefficient decreases from unity according to $|R| \sim  1 - 2 k b  \sim \exp(- 2 k b)$,
where $k$ is the incident wave vector and $b$ a characteristic length which depends on the specifics of the particle surface interaction.
This universal behavior  is a direct result of boundary conditions and continuity of the wave function and its derivative. When considering
scattering from surfaces  which occurs in three dimensional space, $k$ should be replaced by its component perpendicular to the surface.

This quantum threshold reflection phenomenon has been observed for the scattering of ultra cold metastable He atoms on silicon \cite{oberts}
and for rare-gas atoms and small clusters on gratings and surfaces \cite{zhao1,zhao2,wieland1,wieland2} and on a periodic array of
half planes \cite{wieland3}. In this last case, it appears there is a transition from quantum reflection to the regime
where edge diffraction from half planes dominates.
At threshold conditions, where the incident energy is very small, the maximum of the scattered wavefunction is located far away from the
grating/surface  due to the very long de Broglie wave length of the incident particle. This led mistakenly to the idea that quantum reflection takes place
far away from the grating/surface, at distances of typically tens or hundreds of nanometers.
Quantum threshold reflection was claimed to be governed by the long-range attractive van der Waals-Casimir potential
tail which falls off faster than $r^{-2}$. \cite{friedrich1,friedrich2}. It was claimed that the fact that the very weakly bound He$_2$ molecule
which is reflected without dissociating is further proof that only the long range weak forces are at play and these are too weak to dissociate
even such a fragile bond \cite{wieland2}. This would not be the case if the dimer reaches the classical turning point of the interaction potential.

In the semiclassical framework, and within the $k$ linear dependence regime, the semiclassical description of the scattering dynamics
breaks down. Far away from the grating/surface, the long range attractive potential exhibits a region in which the
local de Broglie wavelength is not slowly varying, invalidating a semiclassical description. This occurs  in the so called "badlands" region of the
interaction potential. Quantum reflection was thus associated with this badlands region through a function called "quantality" whose
absolute value displays large deviations from unity in a confined region of the potential, implying that in this region quantum effects are important. \cite{friedrich2,barnea17}
In a series of papers,  we have recently demonstrated theoretically and numerically  \cite{salva1,salva2,salva3} that quantum threshold reflection is determined by the whole
range of the interaction potential. We have observed that the short range region  is also critical for obtaining theoretical
reflection probabilities and  diffraction patterns which are in fairly good agreement with the experimental results of Refs. \cite{zhao1,zhao2,wieland1,wieland2}.
These calculations were carried out by using the elastic close-coupling (CC) formalism \cite{salva4}, which is numerically
exact when convergence is reached. To distinguish between quantum and classical reflection in this type of theoretical calculations, complex
absorbing boundary conditions which prevent the classical reflection from occurring have been employed. \cite{miller,muga}
In Ref. \cite{salva2} we have also shown that the badlands region of the interaction potential is irrelevant in quantum threshold reflection since the wavelength
of the scattering particles at threshold is much longer than the rather small spatial extent of this region.

In the work presented in this paper, we have extended our previous studies to analyze and discuss previously reported coherent reflection of He atom beams from
rough surfaces at grazing incidence. \cite{zhao2} When the component of the incident wave vector of the incoming atom perpendicular to
the surface is very small,
experimental and theoretical reflection probabilities seem to be only dependent on this normal component and approach unity when it vanishes confirming the
corresponding universal behavior. Deviations from linearity are found only at larger values of this normal component. Moreover, we do not find a
transition from quantum to classical reflection  when increasing the normal component of the wave vector, as previously reported
\cite{zhao2}. Finally, the surface roughness at threshold conditions and grazing angles appears to play no important role
in this scattering.

In Section II we review some of the experimental considerations.  The elastic
CC formalism is only briefly outlined in Section III since it has already been described  elsewhere. \cite{salva1,salva3,salva4}
Theoretical results are presented and compared with the experimental ones in Section IV, this is followed by a discussion justifying our
different interpretations and conclusions.

\section{Experimental considerations}

The experiments we want to analyze  have been described in detail  in Refs. \cite{zhao1,zhao2}.
Experimental results have been obtained from a supersonic beam expansion of He atoms at different stagnation temperatures
$T_0 = 300, 50,$ and $8.7$ K corresponding to incidence wave vectors $k$ of 112, 46 and 18 nm$^{-1}$, respectively.
In order to maintain a high atomic flux and narrow velocity distribution  of the incident  beam and avoid cluster formation, different
stagnation pressures $P_0 = 31, 26$ and $0.5$ bar have been used. In the cryogenic free jet expansion of incident
particles, the incident kinetic energy is given by $E_i = (5/2) k_B T_0$ ($k_B$ is Boltzmann's constant).\cite{toennies}
Four types of surfaces have been considered in these experiments \cite{zhao2}: (i) a glass slide which is a simple standard microscope
slide (ISO Norm 8037/I), 1 mm thick and with a surface area of 76 x 26 mm$^2$; (ii) the commercial GaAs wafer which is cut along the
(100) direction and is 50 mm in diameter; (iii) a flat chromium surface of 100 x 30 mm$^2$ area used for comparison with the grating surface
(iv) a 20-$\mu$m-period chromium grating previously used in Ref. \cite{zhao1} (a 56-mm-long microstructured array of 110-nm-thick,
10 $\mu$m-wide and 5-mm-long parallel chromium strips on a flat quartz substrate).
Surfaces are expected to be oxidized or oxygen covered but in spite of this contamination, at grazing angles intense specular reflection peaks
are still observable. Since no information about the roughness of the surfaces employed in the experiments is provided, they are going to be
assumed  flat (except for type (iv)) and well described by one dimension in the $x$-direction.
The incident angle $\theta_i$ is usually varied below 0.1 $mrad$ and up  to 25 $mrad$ and measured with respect to the surface plane.

Angular distributions or diffraction patterns of in-saggital-plane   ($x,z$-plane) scattering are recorded
by rotating the detector and measuring the He signal at each angle. The diffraction angles $\theta_n$ are given by the conservation of the
momentum or Bragg's law
\begin{equation}\label{bragg}
cos \theta_i - cos \theta_n = \frac{n \lambda}{d} = \frac{2 \pi n}{d \, k}
\end{equation}
where $\lambda$ is the de Broglie wave length of the incident particle and the diffraction order is given by $n$. Negative diffraction
orders correspond to diffraction angles close to the surface grating, that is, energy in the perpendicular direction is transferred to the parallel
direction. The specular reflection and non-specular diffraction probabilities are obtained from the integrated intensity of
the reflected peak normalized to the peak area of the incident beam and the total reflection probability is the sum of all of them.
Final results are plotted as a function of the corresponding perpendicular wave vector, that is, along the $z$-direction
\begin{equation}\label{kperp}
k_{perp} \simeq  \frac{\sqrt{5 m k_B  T_0}}{\hbar} \, sin \theta_i
\end{equation}
with $m$ being the He atomic mass.

The main conclusions of  the authors of this experimental work are that at low $k_{perp}$ values the reflection probability is
dominated by quantum threshold reflection, which is manifested by a steep linear decrease of the reflection probability with increasing
$k_{perp}$. At larger $k_{perp}$, the corresponding results start to fan out  and are rationalized in terms of classical reflection from the
inner region of the repulsive interaction potential. The dividing line between both regimes is claimed to be
around $k_{perp}= 0.3$ $nm^{-1}$  and independent of the type of surface studied. Only the fanning out effect is explained in terms of
the surface roughness.

\section{Theory}

The theory employed here has been outlined elsewhere \cite{salva1,salva3} so we will emphasize only the key points: (i) We assume an interaction model with very few free parameters, (ii) The CC equations are solved numerically. The
diffraction probabilities to be compared with the experimental ones are obtained after a fitting procedure of the fundamental potential
parameters and  (iii) The final results are verified to be independent of the absorbing boundary conditions used. This assures that the internal turning
point plays no role in the quantum reflection phenomenon.

A two-dimensional model potential between the incoming particles and the surface is assumed and written as
\begin{equation}
U(x,z) = V(z)\cdot h(x)
\label{Upot}
\end{equation}
where $V(z)$ describes the interaction along the coordinate $z$ perpendicular to the surface and $h(x)$
is the periodic surface along the horizontal coordinate $x$.  For the $z$-direction, the combination of a Morse potential,
$V_M(z)$, at short distances, and an
attractive van der Waals-Casimir tail $V_C$, at large distances, has been shown to be a good description of this interaction
\begin{eqnarray}
 V(z) & = &
  \left\{ \begin{array}{ll}
   V_M(z)=D \left[  e^{- 2 \chi z  } - 2 e^{- \chi z } \right ]
     &, z  < {\bar z} \\
    V_C(z)= - \frac{C_4}{(l + z) z^3}
     &, z \geq  {\bar z} \end{array}
  \right .
\label{vpot}
\end{eqnarray}
Here $C_4 = C_3 l$,  $C_3$ being the van der Waals coefficient  and $l$ a characteristic length which determines  the transition from the van
der Waals ($z \ll l$) to the Casimir($ z \gg l$) regime. The matching point ${\bar z}$ is usually determined by imposing both the continuity
condition for the interaction potential and its first derivative. The range of variation of the $C_3$ parameter is usually known and the
stiffness parameter of the Morse potential $\chi$  is considered a fitting parameter; $D$ is determined from the matching point ${\bar z}$.

The periodic grating function $h(x)$  is described by the so-called unit impulse function and written as
\begin{equation}
h(x) = \sum_{n=- \infty}^{+ \infty} \prod \left ( \frac{x - n d}{a}\right )
\end{equation}
where  $a$ is the width of the strips and $d$ the period  with  $a < d$. The $\prod (y)$-function is the
so-called unit impulse function: $0$ for $|y| > 1/2$, $1$ for $|y| < 1/2$ and $1/2$ for $|y|=1/2$.
In terms of a Fourier series, $h(x)$ is expanded as
\begin{equation}
h(x) =  \sum_{n=- \infty}^{+ \infty} c_n e^{i 2 \pi n x/d}
\label{hfun}
\end{equation}
with $c_0= a/d$, $c_{-n} = c_{n}$ and  $c_n=  (a/d) sinc (n a /d)$,
and $sinc (x) = sin(\pi x)/ \pi x$. When $d = 2 a$ (as in the experimental grating of Ref. \cite{zhao1}),
the terms beyond the sixth order are quite small.
The periodic interaction potential can then be expressed as
\begin{equation}
U(x,z)  =   \sum_{n=- \infty}^{+ \infty} V_n (z) e^{i \frac{2 \pi n x}{d}}
\label{Vfourier}
\end{equation}
where the first term ($n=0$) is the interaction potential $V_0(z) = V(z)$
(see Eq. \ref{vpot}) and the coupling terms ($n \neq 0$) are given by
\begin{equation}
V_n (z) = 2  sinc (n \frac{a}{d}) V(z)
\label{coupling}
\end{equation}
where $d$ and $a$ are the period and width of the strips.

As  has been recently shown \cite{salva1,salva3}, the elastic scattering of the incident
particles with the surface is theoretically well described by the CC formalism. The corresponding CC differential equations are
written as
\begin{equation}\label{close_coupling}
 \left[ \frac{\hbar^2}{2m} \frac{d^2}{dz^2} + \frac{\hbar^2}{2m} k_{n,z}^2
- V_{0}(z) \right] \psi_{n}(z) ~=~ \sum\limits_{n \neq n'}
V_{n - n'}(z) \psi_{n'}(z)
\end{equation}
with  $\frac{\hbar^2}{2m} k_{n,z}^2$  being the z-component of the kinetic energy of the
scattered particles.  The square z-component of the wave vector is given by
\begin{equation}\label{tr}
 k_{n,z}^2 ~=~ k_i^2 - \left( k_i \sin \theta_i + \frac{2 \pi n}{d} \right)^2.
\end{equation}
with $\theta_i$ measured with respect to the normal to the surface. This theoretical angle is complementary to the experimental incident angle.
Thus, when comparing with experimental results, theoretical positive $n$ diffraction orders
correspond to experimental negative  ones. The effective potentials labelled by $n$
$V_{0}(z)+ \frac{\hbar^2}{2m} (k_i \sin \theta_i + 2 \pi  n / d)^2$ in Eq.(\ref{close_coupling})
represent diffraction channels. The asymptotic energies depend on $n$ and  the incident energy and polar angle.
As is known, open (closed) diffraction channels have a positive (negative) normal kinetic energy
$\hbar^2 k_{n,z}^2/(2m) $. The coupling between channels $V_{n - n'}(z)$ is given by Eq. (\ref{coupling}) since $n-n'$ is always an
integer number. The diffraction intensities or reflection probabilities, obtained by solving the CC equations given by Eq. (\ref{close_coupling})
with the usual boundary conditions \cite{salva4}, are expressed as
\begin{equation}
I_n = |S_{n0}|^2
\end{equation}
where $S_{nn'}$ are the elements of the unitary scattering matrix. Their square absolute values give the diffraction intensity or probability
for an incident wave at the specular channel ($n' =0$) and exiting by any of the open diffraction channels labelled by $n$.

\begin{figure}
	\includegraphics[scale=0.4,angle=0]{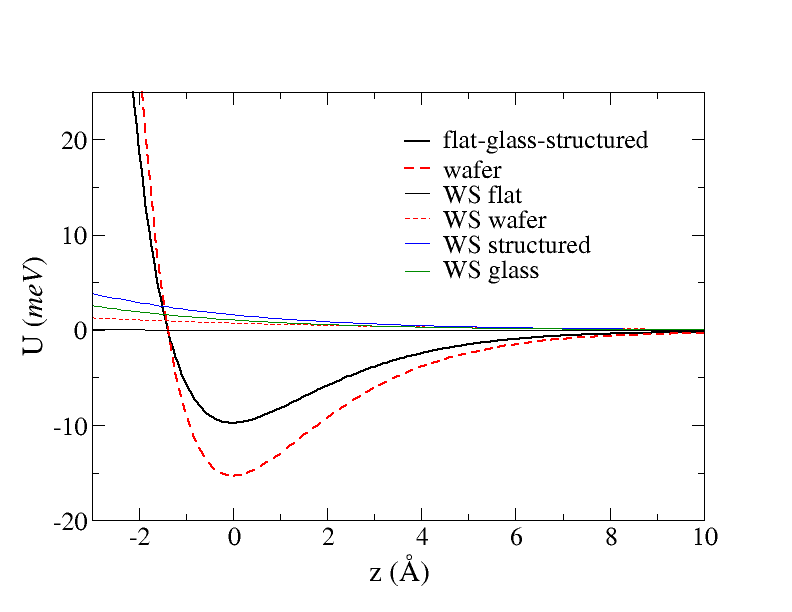}
	
	\caption{(Color online)  The interaction potentials for the glass slide, flat and structured Cr surfaces (solid curve) and wafer (dashed curve)
are plotted. The WS potentials are also plotted for all the surfaces used in each case (gray curves). }
	
	\label{pot-surface}
\end{figure}

As mentioned above, the interaction potential given by Eq. (\ref{vpot}) displays classical turning points due to the
repulsive part of the Morse potential. To distinguish between  quantum threshold reflection  and classical reflection from the inner repulsive
part of the Morse potential,
absorbing boundary conditions have to be imposed \cite{miller,muga} in the inner part.  A Woods-Saxon (WS) potential is  added to the
imaginary part of the diffraction channel potentials
\begin{equation}
V_{WS} = \frac{A}{1 + e^{\alpha  \chi (z-z_i)}}
\label{17}
\end{equation}
which is essentially zero in the physically relevant interaction region and turns on sufficiently rapidly but smoothly at the left edge of the numerical
grid for the integration to absorb the flux. The fitting parameters  of this WS potential are $A$ and $\alpha$. The resulting scattering matrix
${\bar S}$ is then no longer unitary. The diffraction intensities are  given by ${\bar I}_n = |{\bar S}_{n0}|^2$ and the total quantum reflection
probability is calculated as
\begin{equation}\label{qrp}
P^{QR} = \sum_n |{\bar S}_{n0}|^2   < 1
\end{equation}
for each initial condition. Due to the absorbing potential the theoretical diffraction efficiencies are defined as the ratio of the diffraction intensity
${\bar I}_n$ to the total quantum reflection probability $P^{QR}$ rather than to the total incident flux in order to compare to the experimental
results. In Figure \ref{pot-surface}, the potentials for the glass slide, flat and structured Cr surfaces are plotted together with the
WS potentials used in each case. For flat surfaces $h(x)=1$ and no diffraction channels are present, only specular reflection is present.
The WS potential is  then added to the specular channel.

%
%
\begin{table}
	\caption{Parameters of the interaction potential $V(z)$ for the four surfaces. The stiffness parameter of the Morse potential, $\chi$, is a
		free parameter fitted to reproduce the corresponding experimental results and $D$ is the well depth. The characteristic lengths $l$ and
		parameters $C_3$ are based on values reported in previous works \cite{zhao2}.}
	\vspace{1cm}
	\centering
	\begin{tabular}{|c||c||c||c||c|}
		\hline       Parameters & Glass slide & GaAs wafer &  Flat Cr & Structured Cr   \\
		\hline      $\chi$ ($\AA^{-1}$) &  0.5 & 0.5 & 0.5 & 0.5 \\
		\hline      D (meV) & 9.8 & 15.3 & 9.8 & 9.8  \\
		\hline      l ($\AA$ ) & 93 & 93 & 93 & 93  \\
		\hline       C$_3$ (10$^{-50}$ J m$^3$) & 3.5  & 5.5 & 3.5 & 3.5  \\ \hline
	\end{tabular}
	\label{table1}
	\vspace{1cm}
\end{table}

\section{ Results and discussion}

The parameters used in the elastic CC calculations for the  perpendicular
potential are  displayed in Table I.  As previously noted, the only real fitting parameter is $\chi$ since the values of $C_3$ have been taken from
Ref. \cite{zhao2} and only slightly modified. The values obtained from  fitting to the experimental results when solving numerically
the one-dimensional Schr\"odinger equation only for the attractive potential were 3.5, 5.5, 3.5 and 3.5 $\times 10^{-50}$ $J m^3$ 
for the glass slide,
wafer, flat and structured Cr, respectively. The well depth
is related to $\chi$ according to our procedure to evaluate the matching point $\bar z$.  A characteristic length $l$ of 9.3 nm for He for the
transition from the Casimir to the van der Waals regimes in the long range attractive potential is a well
accepted value in the literature. \cite{zhao2} For the numerical integration, the range of distances in the vertical $z$-axis was taken to be -13 to
2,000 $\AA$.  In this way the wave function was forced to vanish at $z=-13$  $\AA$ ($z=0$ corresponds to the location of
the minimum of the attractive Morse potential).

The parameters of the
WS potential were varied with the incident wave vector. The  detailed procedure for using these complex absorbing boundary conditions
for the diffraction channels has been described elsewhere. \cite{salva1,salva3} Due to the fact that no information is available on the structure
of the experimental surfaces used, we have assumed that they are flat, except for the structured Cr surface. Thus, $h(x)=1$ is chosen for the
glass slide, GaAs wafer and flat Cr surfaces leading to only specular reflection or reflectivity.
The characteristics of the grating Cr surface are the same as reported in Ref. \cite{zhao1} with $h(x)$ given by Eq. (\ref{hfun}).

\begin{figure}
	\includegraphics[scale=0.4,angle=0]{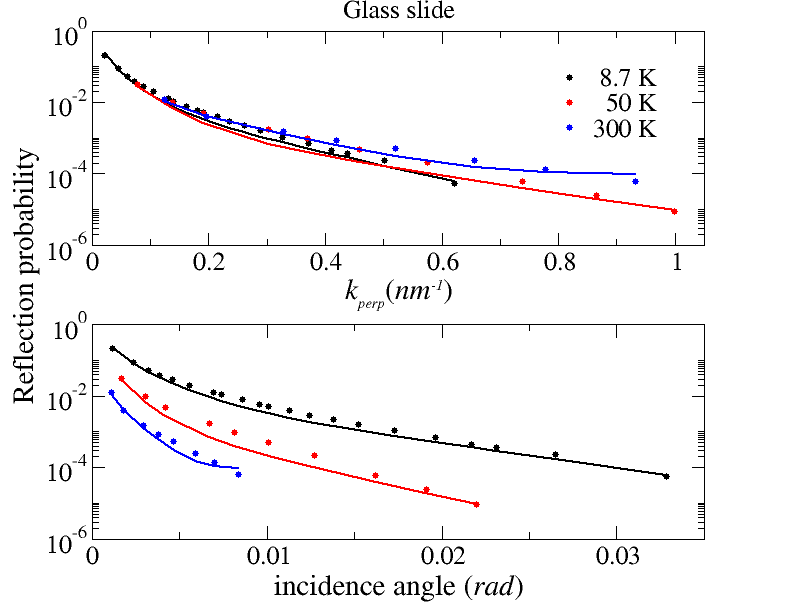}
	
	\caption{(Color online)  Reflection probabilities for He atoms scattering from a glass slide at three different source temperatures 8.7 K
		(black labels), 50 K (red labels, light gray ) and 300 K (blue labels, dark gray). In the top panel, probabilities are plotted as a function
        of $k_{perp}$ in $nm^{-1}$ whereas in the bottom panel, they are plotted versus the incident angle in $rad$. Points are the
        experimental results and solid lines show  the results of the present elastic CC computations.}
	
	\label{glass-slide}
\end{figure}

A measure of the quality of the reflection probability fits for each surface can be given by
the square root of the relative deviation $\sigma$  which is defined as
\begin{equation}\label{sigma}
\sigma =  \sqrt{   \frac{1}{N (N-1)}  \sum_{j=1}^N \left| \frac{P_j^{exp}- P_j^{theo}}{P_j^{exp}} \right|^2}  .
\end{equation}
Here, $N$ is the total number of initial conditions given by the $k_{perp}$  points for each surface and $j=1, \cdots , N$.
The smaller the sigma coefficient, the better the quality of the fit.

Reflection probabilities for He atoms scattering from a glass slide at three different source temperatures of 8.7 K
(black labels), 50 K (red labels, light gray) and 300 K (blue labels, dark gray) are plotted in Figure \ref{glass-slide} as a function
of $k_{perp}$ in
$nm^{-1}$ (top panel) and versus the incident
angle in $rad$ (bottom panel). Points are the experimental results and solid curves our elastic CC results. The overall agreement is fairly good.
The universal linear behavior of the quantum reflection probability in the top panel is observed  for $k_{perp}$ values  below
0.2  $nm^{-1}$, depending on
the stagnation temperature. After this small region of $k_{perp}$, one finds  that the value of  $k_{perp}$ causes  this probability
to fan out. The same
is observed  in the bottom panel  where the incident angle also contributes to this effect.  The quantum threshold reflection clearly decreases
with $k$ and $\theta_i$.  The $\sigma$ coefficient is 0.063  for 8.7 K, 0.143 for 50 K and 0.114  for 300 K.

Whereas the same observations are extracted from
analyzing the experimental and theoretical results, our interpretation is different. We attribute our results to only quantum threshold reflection,
not  to classical reflection from the inner region of the potential even at higher $k_{perp}$. Our theoretical calculations are
preventing classical reflection due to the use of the
complex absorbing boundary conditions. The classical turning points of the perpendicular potential are not reached. The universal
linear behavior is gradually lost with increasing  $k_{perp}$. 
The transition from quantum to classical reflection cannot be observed by varying the incident wave
vector.  Moreover, from our calculations, it becomes clear that  the fanning out effect observed  for this surface should not be attributed
to the surface
roughness as mentioned in Ref. \cite{zhao2} since we used a flat surface.

\begin{figure}
		\includegraphics[scale=0.4,angle=0]{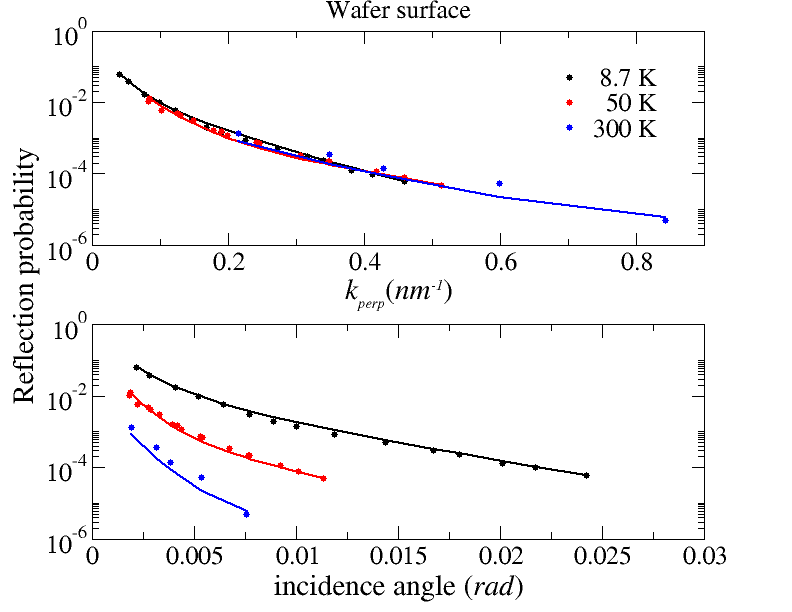}
	\caption{(Color online) The same as in Figure (\ref{glass-slide})  but for the GaAs wafer.}
	
	\label{GaAs-wafer}
\end{figure}

The same arguments and conclusions  are valid also for
the GasAs wafer surface whose results are plotted in Figure \ref{GaAs-wafer}. The degree of this fanning out is the smallest for the wafer
as compared to the three other surfaces studied.
The hierarchy of surface roughness determined by qualitative AFM measurement \cite{zhao2} indicates that the root-mean-square
surface roughness is smallest for the  chromium surface. In particular, the glass slide is larger than for the wafer. It is thus suggestive that
the extent of surface roughness affects the extent of fanning out at higher $k_{perp}$. However, from our computations we conclude that
this roughness has at most a minor effect on the quantum threshold reflection. The $\sigma$ coefficient is 0.044  for 8.7 K,
0.049  for 50 K and 0.215  for 300 K.

\begin{figure}
	\includegraphics[scale=0.4,angle=0]{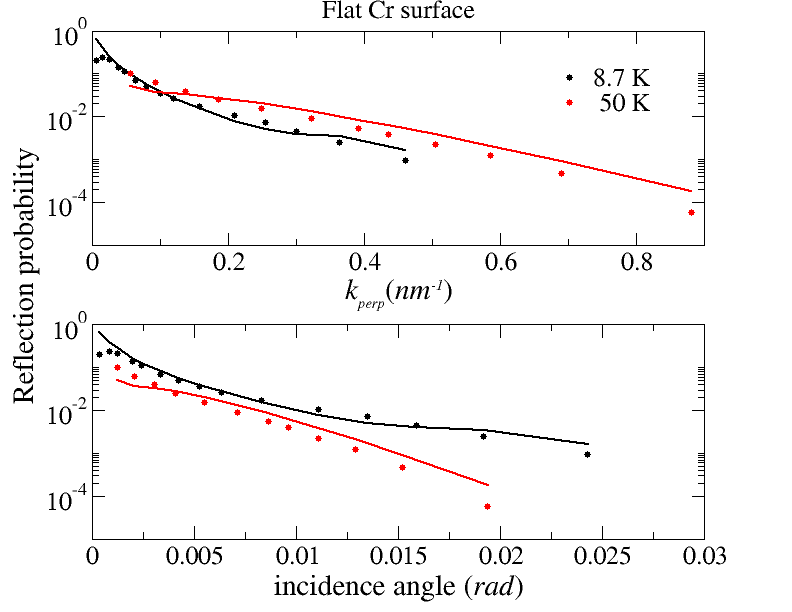}
	\caption{(Color online) The same as in Figure (\ref{glass-slide})  but for a flat Cr surface. }
	
	\label{Cr-flat}
\end{figure}

\begin{figure}
	\includegraphics[scale=0.4,angle=0]{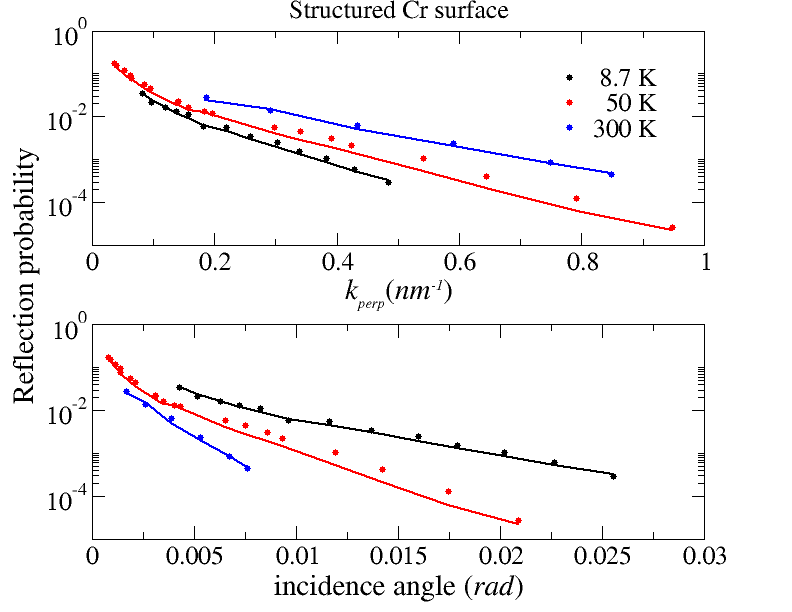}
	\caption{(Color online)  The same as in Figure (\ref{glass-slide})  but for the structured Cr surface.}
		
		\label{Cr-structured}
	\end{figure}

The next two surfaces are interesting to study to see what is the real effect of the roughness. In Figures \ref{Cr-flat} and \ref{Cr-structured}, we
present the quantum reflection probabilities for a flat and structured chromium surface, respectively. The linear behavior seems to take place
for perpendicular wave vectors less than 0.2 nm$^{-1}$ in both cases. The fanning out effect is also more pronounced in the structured surface.
However, the conclusions are the same as previously mentioned for the other two surfaces. The $\sigma$ coefficient is 0.09
for 8.7 K, 0.24 for 50 K for the flat surface, whereas for the structured surface we have  0.04  for 8.7 K, 0.063  for 50 K and
0.058 for 300 K.

\section{Concluding remarks}	
	
In this work, we presented a theoretical study, based on the closed coupled equations method, of the scattering of He atom beams at grazing angles
and threshold conditions on four surfaces; a glass slide, a GaAs wafer, and a flat and structured chromium surface. The reasonable agreement
between our results and the experimental ones indicates the reflection observed in the experiment should be attributed solely to quantum
threshold reflection and not to classical reflection from the repulsive inner region of the potential.  For all
the surfaces studied we also observe the fanning out effect of the reflection probability with increasing incident energy but attribute this
to the quantum threshold reflection rather than the classical reflection from the inner turning point.  The universal linear dependence of the
reflection probability on the perpendicular component of the incident wave vector is gradually lost.
Finally, the
decrease of the quantum threshold reflection probability with the incident wave vector and incident angle is found to be less pronounced for the
wafer  and much more for the structured Cr surface due to its roughness. The present computational result again indicates  that the whole
interaction potential is needed to correctly describe the quantum threshold reflection phenomenon, not only qualitatively but also quantitatively.

\vspace{1cm}

{\bf Acknowledgment}: The authors would like to thank W. Sch\"ollkopf and B. S. Zhao for
providing us with their experimental results. This work is supported by the Programa Nacional de Ciencias Básicas de Cuba
PNCB: P223LH001-108 (G.R.L. and J.R.S.), by a grant with Ref. FIS2017-83473-C2-1-P from the
Ministerio de Ciencia, Innovaci\'on y Universidad (Spain) (S.M.A.) and by grants of the Israeli Science Foundation  and the Minerva Foundation,
Munich. (E.P.).


\end{document}